\documentclass{sig-alternate}

\usepackage{mathrsfs}
\usepackage{graphicx}
\usepackage{bbm}
\usepackage{amsmath}
\usepackage{alltt, amssymb}

\def\Q {\ensuremath{\mathbb{Q}}}
\def\N {\mathbb{N}}

\def\F {\ensuremath{\mathbf{F}}}

\def\K {\ensuremath{\mathbb{K}}}

\def\A {\ensuremath{\mathsf{A}}}

\def\M{\ensuremath{\mathsf{M}}}
\def\T{\ensuremath{\mathsf{T}}}

\def\val {\ensuremath{{\rm val}}}
\def\ct {\ensuremath{{\rm lc}}}
\def\tt {\ensuremath{{\rm lt}}}
\def\op {\ensuremath{\mathbf{O}}}
\def\o {\ensuremath{\mathsf{o}}}

\def\mymod {\ensuremath{\mathsf{mod}}}

\def\dom {\ensuremath{\mathsf{dom}}}
\def\g {\ensuremath{\mathsf{G}}}

\def\Lg {\ensuremath{\mathsf{Log}}}
\def\lg {\ensuremath{\mathsf{L}}}

\def\Xp {\ensuremath{\mathsf{Exp}}}
\def\xp {\ensuremath{\mathsf{E}}}

\def\shift {\ensuremath{\mathsf{A}}}
\def\Shift {\ensuremath{\mathsf{Shift}}}

\def\scale {\ensuremath{\mathsf{M}}}
\def\Scale {\ensuremath{\mathsf{Scale}}}

\def\power {\ensuremath{\mathsf{P}}}
\def\Power {\ensuremath{\mathsf{Power}}}

\def\root {\ensuremath{\mathsf{R}}}
\def\Root {\ensuremath{\mathsf{Split}}}

\def\LinComb {\ensuremath{\mathsf{Comb}}}            
\def\Diag{\ensuremath{\mathsf{\Delta}}}

\def\Rev {\ensuremath{\mathsf{Rev}}}
\def\inv {\ensuremath{\mathsf{Inv}}}

\def\mul {\ensuremath{\mathsf{Mul}}}
\def\Eval {\ensuremath{\mathsf{Eval}}}

\def\myproof{\noindent{\sc Proof.}~}
\def\foorp{\hfill$\square$}

\newtheorem{Def}{Definition}
\newtheorem{theorem}{Theorem}
\newtheorem{Coro}{Corollary}
\newtheorem{Prop}{Proposition}

\newtheorem{Lemma}{Lemma}

\title{Power Series Composition and Change of Basis}
\numberofauthors{3}
\author{
\alignauthor Alin Bostan\\
\affaddr{Algorithms Project}\\
\affaddr{INRIA Rocquencourt}\\ 
\affaddr{France}\\
\affaddr{Alin.Bostan@inria.fr} 
\alignauthor Bruno Salvy\\
\affaddr{Algorithms Project}\\
\affaddr{INRIA Rocquencourt}\\ 
\affaddr{France}\\
\affaddr{Bruno.Salvy@inria.fr} 
\alignauthor {\'E}ric Schost\\
\affaddr{ORCCA and CSD}\\
\affaddr{University of Western Ontario}\\
\affaddr{London, ON, Canada}\\
\affaddr{eschost@uwo.ca} 
}

\begin{document}

\conferenceinfo{ISSAC'08,} {July 20--23, 2008, Hagenberg, Austria.}  
\CopyrightYear{2008} 
\crdata{978-1-59593-904-3/08/07}

\maketitle
\begin{abstract}
Efficient algorithms are known for many operations on truncated power
series (multiplication, powering, exponential, \dots). Composition is
a more complex task. We isolate a large class of power series for
which composition can be performed efficiently. We deduce fast
algorithms for converting polynomials between various bases, including
Euler, Bernoulli, Fibonacci, and the orthogonal Laguerre, Hermite,
Jacobi, Krawtchouk, Meixner and Meixner-Pollaczek.
\end{abstract}


\vspace{1mm}
 \noindent
 {\bf Categories and Subject Descriptors:} \\
\noindent I.1.2 [{\bf Computing Methodologies}]:{~} Symbolic and Algebraic
  Manipulation -- \emph{Algebraic Algorithms}
 
 \vspace{1mm}
 \noindent
 {\bf General Terms:} Algorithms, Theory
 
 \vspace{1mm}
 \noindent
 {\bf Keywords:} Fast algorithms, transposed algorithms, basis conversion, orthogonal polynomials.


\section{Introduction}\label{sec:intro}

Through the Fast Fourier Transform, fast polynomial multiplication has
been the key to devising efficient algorithms for polynomials and
power series. Using techniques such as Newton iteration or
divide-and-conquer, many problems have received satisfactory
solutions: polynomial evaluation and interpolation, power series
exponentiation, logarithm, \dots~can be performed in quasi-linear
time.

In this article, we discuss two questions for which such fast
algorithms are not known: power series composition and change of basis
for polynomials. We isolate special cases, including most
common families of orthogonal polynomials, for which our algorithms
reach quasi-optimal complexity.

\smallskip\noindent{\bf Composition.} Given a power series $g$ with
coefficients in a field $\K$, we first consider the map of
evaluation at $g$
$$\Eval_{m,n}(.,g): A \in \K[x]_m \mapsto A(g) \bmod x^n \in \K[x]_n.$$
Here, $\K[x]_m$ is the $m$-dimensional $\K$-vector space of
polynomials of degree less than $m$. We note~$\Eval_n$ for 
$\Eval_{n,n}$.

To study this problem, as usual, we denote by $\M$ a
\emph{multiplication time} function, such that polynomials of degree
less than $n$ can be multiplied in $\M(n)$ operations in~$\K$. We
impose the usual super-linearity conditions
of~\cite[Chap.~8]{GaGe99}. Using Fast Fourier Transform algorithms,
$\M(n)$ can be taken in $O(n \log(n))$ over fields with suitable roots
of unity, and $O(n \log(n)\log\log(n))$ in
general~\cite{ScSt71,CaKa91}.

If $g(0)=0$, the best known algorithm, due to Brent and Kung,
uses~$O(\sqrt{n\log n} \, \M(n))$ operations in~$\K$~\cite{BrKu78}; in
small characteristic, a quasi-linear algorithm is
known~\cite{Bernstein98b}. There are however special cases of power
series~$g$ with faster algorithms: evaluation at $g=\lambda x$ takes
linear time; evaluation at $g=x^k$ requires \emph{no} arithmetic
operation. A non-trivial example is $g=x+a$, which takes time
$O(\M(n))$ when the base field has characteristic zero or large
enough~\cite{AhStUl75}. Brent and Kung~\cite{BrKu78} also showed how
to obtain a cost in $O(\M(n)\log(n))$ when $g$ is a polynomial; this
was extended by van der Hoeven~\cite{Hoeven02} to the case where $g$
is algebraic over $\K(x)$. In \S\ref{sec:compose}, we prove that
evaluation at $g=\exp(x)-1$ and at $g=\log(1+x)$ can also be performed
in $O(\M(n) \log(n))$ operations over fields of characteristic zero or
larger than~$n$.

Using associativity of composition and the linearity of the
map~$\Eval_{m,n}$, we show in \S\ref{sec:compose} how to use these
special cases as building blocks, to obtain fast evaluation algorithms
for a large class of power series.  This idea was first used by
Pan~\cite{Pan98}, who applied it to functions of the form
$(ax+b)/(cx+d)$. Our extensions cover further examples
such as~$2x/(1+x)^2$ or $(1-\sqrt{1-x^2})/x$, for which we improve 
the previously known costs.

\smallskip\noindent{\bf Bivariate problems.}  Our results on the cost
of evaluation (and of the transposed operation) are applied in
\S\ref{sec:basis} to special cases of a more general composition,
reminiscent of umbral operations~\cite{Roman05}. Given a
\emph{bivariate} power series $\F=\sum_{j\ge0}\xi_j(x)t^j$, we
consider the linear map
\[\Eval_n(.,\F,t):  (a_0,\dots,a_{n-1})  \mapsto  \sum_{j < n}\xi_j(x) a_j \bmod x^n.\]
For instance, with $$\F= \frac1{1-tg(x)}=\sum_{j\ge 0} g(x)^jt^j,$$
this is the map~$\Eval_n(.,g)$ seen before.  For general $\F$, the
conversion takes quadratic time (one needs $n^2$ coefficients for
$\F$). Hence, better algorithms can only been found for structured
cases; in \S\ref{sec:basis}, we isolate a large family of bivariate
series~$\F$ for which we can provide such fast algorithms.
This approach follows Frumkin's~\cite{Frumkin95}, which was specific
to Legendre polynomials.

\smallskip\noindent{\bf Change of basis.} Our framework captures in
particular the generating series of many classical polynomial
families, for which it yields at once conversion algorithms between
the monomial and polynomial bases, in both directions.

Thus, we obtain in \S\ref{sec:applications} change of basis
algorithms of cost only $O(\M(n))$ for all of Jacobi, Laguerre and
Hermite orthogonal polynomials, as well as Euler, Bernoulli, and Mott
polynomials (see Table~\ref{Fig:M1}). These algorithms are derived
in a uniform manner from our composition algorithms; they improve upon
the existing results, of cost $O(\M(n) \log(n))$ or $O(\M(n)
\log^2(n))$ at best (see below for historical comments).

We also obtain $O(\M(n) \log(n))$ conversion algorithms for a large
class of Sheffer sequences~\cite[Chap.~2]{Roman05}, including
actuarial polynomials, Poisson-Charlier polynomials and Meix\-ner
polynomials (see Table~\ref{Fig:M2}).

\smallskip\noindent{\bf Transposition.} A key aspect of our results is
their heavy use of transposed algorithms. Introduced under this name
by Kaltofen and Shoup, the \emph{transposition principle} is an
algorithmic theorem with the following content: given an algorithm
that performs an $r \times s$ matrix-vector product $b \mapsto M b$,
one can deduce an algorithm with the same complexity, up to $O(r+s)$
operations, and that performs the transposed matrix-vector product $c
\mapsto M^t c$. In other words, this relates the cost of computing a
$\K$-linear map $f:\ V \to W$ to that of computing the transposed map
$f^t:\ W^* \to V^*$.

For the transposition principle to apply, some restrictions must be
imposed on the computational model: we require that only linear
operations in the coefficients of $b$ are performed (all our
algorithms satisfy this assumption). See~\cite{BuClSh97} for a precise
statement, Kaltofen's ``open problem''~\cite{Kaltofen00} for further
comments and~\cite{BoLeSc03} for a systematic review of some classical
algorithms from this viewpoint.

To make the design of transposed algorithms transparent, we choose as
much as possible to describe our algorithms in a ``functional''
manner. Most of our questions boil down to computing linear maps
$\K[x]_m \to \K[x]_n$; expressing algorithms as a factorization of
these maps into simpler ones makes their transposition
straightforward. In particular, this leads us to systematically
indicate the dimensions of the source (and often target) space as
a subscript.

\smallskip\noindent{\bf Previous work.} The question of efficient
change of basis has naturally attracted a lot of attention, so that
fast algorithms are already known in many cases.

Gerhard~\cite{Gerhard00} provides $O(\M(n) \log(n))$ conversion
algori\-thms between the falling factorial basis and the monomial basis:
we recover this as a special case.  The general case of Newton
interpolation is discussed in~\cite[p. 67]{BiPa94} and
developed in~\cite{BoSc05}. The algorithms have cost $O(\M(n)
\log(n))$ as well.

More generally, if $(P_i)$ is a sequence of polynomials satisfying a
recurrence relation of fixed order (such as an orthogonal family), the
conversion from $(P_i)$ to the monomial basis $(x^i)$ can also be
computed in $O(\M(n) \log(n))$ operations: an algorithm is given
in~\cite{PoStTa98}, and an algorithm for the transpose problem is
in~\cite{DrHeRo97}. Both operate on real or complex arguments, but the
ideas extend to more general situations.  Alternative algorithms,
based on structured matrices techniques, are given
in~\cite{Heinig01}. They perform conversions in both directions in
cost $O(\M(n) \log^2(n))$.

The overlap with our results is only partial: not all families
satisfying a fixed order recurrence relation fit in our framework;
conversely, our method applies to families which do not necessarily
satisfy such recurrences (the work-in-progress~\cite{BoSaSc08}
specifically addresses conversion algorithms for orthogonal polynomials).

Besides, special algorithms are known for converting between
particular families, such as Chebyshev, Legendre and
B\'ezier~\cite{LiZh98,BaPe04}, with however a quadratic
cost. Floating-point algorithms are known as well, of cost~$O(n)$ for
conversion from Legendre to Chebyshev bases~\cite{AlRo91} and $O(n
\log(n))$ for conversions between Gegenbauer bases~\cite{Keiner07},
but the results are approximate. Approximate conversions for the
Hermite basis are discussed in~\cite{LeRoCh07}, with cost $O(\M(n)
\log(n))$.

\smallskip\noindent{\bf Note on the base field.} For the sake of
simplicity, in all that follows, the base field is supposed to have
characteristic~0. All results actually hold more generally, for fields
whose characteristic is sufficiently large with respect to the target
precision of the computation. However, completely explicit estimates
would make our statements cumbersome.


\section{Composition}\label{sec:compose}
Associativity of composition can be read both ways: in the identity
$A(f\circ g)=A(f)\circ g$, $f$ is either composed on the left of~$g$
or on the right of~$A$.  In this section, we discuss the consequences
of this remark.  We first isolate a class of operators~$f$ for which
both left and right composition can be computed fast. Most results are
known; we introduce two new ones, regarding exponentials and logarithms.
Using these as building blocks, we then define \emph{composition
sequences}, which enable us to obtain more complex functions by
iterated compositions. We finally discuss the cost of the map
$\Eval_{n}$ and of its inverse for such functions, showing how to
reduce it to $O(\M(n))$ or $O(\M(n)\log(n))$.


\subsection{Basic Subroutines}\label{ssec:basic}
We now describe a few basic subroutines that are the building
blocks in the rest of this article.
   
\smallskip\noindent{\bf Left operations on power series.}  In
Table~\ref{tab:leftcomp}, we list basic composition operators,
defined on various subsets of $\K[[x]]$.
Explicitly, any such operator $\o$ is defined on a \emph{domain}
$\dom(\o)$, given in the third column. Its action on a power
series~$g\in\dom(\o)$ is given in the second column, and the cost of
computing~$\o(g)\bmod x^n$ is given in the last column. 

\begin{table}[!!!h]
\centering
\begin{tabular}{l@{\hspace{-0em}}c@{\hspace{-0em}}c@{\hspace{0em}}c}
{\hspace{-0.8em}}Operator&Action&Domain&Cost\\
\hline
{\hspace{-0.8em}}$\A_a$ (add)&$a+g$&$\K[[x]]$&1\\
{\hspace{-0.8em}}$\M_\lambda$ (mul)&$\lambda g$&$\K[[x]]$&$n$\\
{\hspace{-0.8em}}$\power_k$ (power)&$g^k$&$\K[[x]]$&\hspace{-1ex}$O(\log k+\M(n))$\\  
{\hspace{-0.8em}}$\root_{k,\alpha,r}$ (root)&$g^{1/k}$&{~}$\alpha^kx^{rk}(1+x\K[[x]])$&$O(\M(n))$\\
{\hspace{-0.8em}}$\inv$ (inverse)&$1/g$&$\K^*+x\K[[x]]$&$O(\M(n))$\\
{\hspace{-0.8em}}$\xp$ (exp.)&{~}$\exp(g)-1$&$x\K[[x]]$&$O(\M(n))$\\
{\hspace{-0.8em}}$\lg$ (log.)&$\log(1+g)$&$x\K[[x]]$&$O(\M(n))$\\       
\hline
\end{tabular}
\caption{Basic Operations on Power Series\label{tab:leftcomp}}
\end{table}

Some comments are in order. For addition and multiplication, we take
$a \in \K$ and $\lambda$ in $\K^*$. To lift indeterminacies, the value
of $\root_{k,\alpha,r}(g)$ is defined as the unique power series with
leading term~$\alpha x^r$ whose~$k$th power is~$g$; observe that to
compute $\root_{k,\alpha,r}(g) \bmod x^n$, we need $g$ modulo
$x^{n+r(k-1)}$ as input. Finally, we choose to subtract~1 to the
exponential so as to make it the inverse of the logarithm.
All complexity results are known; they are obtained by Newton
iteration~\cite{Brent75}.

\smallskip\noindent{\bf Right operations on polynomials.}  In
Table~\ref{tab:rightcomp}, we describe a few basic linear maps
on~$\K[x]_m$ (observe that the dimension $m$ of the source is
mentioned as a subscript). Their action on a polynomial
\[A(x)=a_0+\dots+a_{m-1}x^{m-1}\in\K[x]_m\]
is described in the third column. In the case of powering, it is
assumed that~$k\in\N_{>0}$. Here and in what follows, we  freely
identify $\K[x]_m$ and $\K^m$, through the isomorphism
$$ \sum_{i< m} a_i x^i  \in \K[x]_m \ \leftrightarrow \ (a_0,\dots,a_{m-1}) \in \K^m.$$

\begin{table}[!!!t]
\centering
\begin{tabular}{l@{\hspace{0.5em}}c@{\hspace{0.5em}}c@{\hspace{0.5em}}c}
{\hspace{-0.8em}}Name&Notation&Action&Cost\\ 
\hline
{\hspace{-0.8em}}Powering&$\Power_{m,k}$&$A(x^k)$&0\\
{\hspace{-0.8em}}Reversal&$\Rev_m$&$x^{m-1}A(1/x)$&0\\
{\hspace{-0.8em}}Mod&$\mymod_{m,n}$&$A \bmod x^n$&0\\
{\hspace{-0.8em}}Scale&$\Scale_{\lambda,m}$&$A(\lambda x)$&$O(m)$\\
{\hspace{-0.8em}}Diagonal&$\Diag_m(.,s_i)$&$\sum{a_is_ix^i}$&$m$\\
{\hspace{-0.8em}}Multiply&~$\mul_{m,n}(.,P)$~&~$AP\bmod x^n$&$~\M(\max(n,m))$\\
{\hspace{-0.8em}}Shift&$\Shift_{a,m}$&$A(x+a)$&~$\M(m)+O(m)$\\
\hline
\end{tabular}
\caption{Basic Operations on Polynomials\label{tab:rightcomp}}
\vspace{-2ex}
\end{table} 

\noindent All of the cost estimates are straightforward, except for
the shift, which, in characteristic~0, can be deduced from the other
ones by the factorization~\cite{AhStUl75}:
\[\Shift_{a,m}=\Diag_m(\Rev_m(\mul_{m,m}(\Rev_m(\Diag_m(.,i!)),P)),1/i!),\]
where $P$ is the polynomial $\sum_{i=0}^{n-1} a^i x^i/i!$.
We continue with some equally simple operators, whose description
however requires some more detail. For $k \in \N_{>0}$, any polynomial $A$
in $\K[x]$ can be uniquely written as
$$A(x)=A_{0/k}(x^k) + A_{1/k}(x^k)x + \cdots + A_{k-1/k}(x^k) x^{k-1}.$$
Inspecting degrees, one sees that if $A$ is in $\K[x]_m$, then $A_{i/k}$ is in $\K[x]_{m_i}$, with    
\begin{equation}\label{eq:ni}
m_i=\lfloor{m/k}\rfloor+\begin{cases}1&\text{if $i\le m \bmod k$,}\\
0&\text{otherwise.}\end{cases}
\end{equation}
This leads us to define the map $\Root_{m,k}:$ 
$$ A \in \K[x]_m \mapsto (A_{0/k},\dots,A_{k-1/k}) \in \K[x]_{m_0}
\times \cdots \times \K[x]_{m_{k-1}}.$$ It uses no arithmetic operation.
We also use linear combination with polynomial coefficients.
Given polynomials $G_0,\dots,G_{k-1}$ in $\K[x]_m$, we
denote by
$$\LinComb_{m}(.,G_0,\dots,G_{k-1}):\ \  \K[x]_m^k \to \K[x]_m$$ the map sending
$(A_0,\dots,A_{k-1}) \in \K[x]_m^k$ to
$$A_0 G_0 + \cdots + A_{k-1} G_{k-1} \bmod x^{m} \in \K[x]_m.$$
It can be computed in~$O(k\M(m))$ operations.
Finally, we extend our set of subroutines on polynomials with the
following new results on the evaluation at $\exp(x)-1$ and $\log(1+x)$.
\begin{Prop}\label{Prop:xp} The maps
\begin{align*}
\Xp_{m,n}: A \in \K[x]_m &\mapsto A(\exp(x)-1) \bmod x^n \in \K[x]_{n},\\
\Lg_{m,n}: A \in \K[x]_m &\mapsto A(\log(1+x)) \bmod x^n \in \K[x]_{n}
\end{align*}
can be computed in $O(\M(n)\log(n))$ arithmetic operations.  
\end{Prop}
\myproof We start by truncating $A$ modulo $x^n$, since
$$\Xp_{m,n}(A) = \Xp_{m,n}(A \bmod x^n).$$ After shifting by $-1$, we
are left with the question of evaluating a polynomial in $\K[x]_n$ at
$\sum_{i < n} x^i/i!$. Writing its matrix shows that this map factors
as $\Diag_n({\sf MultiEval}^t_n(.),1/i!),$ where ${\sf MultiEval}_n$
is the map
$$A \in \K[x]_n \mapsto (A(0),\dots,A(n-1))\in \K^n.$$ To summarize,
we have obtained that
$$\Xp_{m,n}(A) = \Diag_n({\sf
MultiEval}^t_n(\Shift_{-1,n}(\mymod_{m,n}(A))),1/i!).$$ Using fast
transposed evaluation~\cite{CaKaLa89,BoLeSc03}, $\Xp_{m,n}(A)$ can
thus be computed in $O(\M(n)\log(n))$ operations.  Inverting these
computations leads to the factorization
$$\Lg_{m,n}(A) = \Shift_{1,n}({\sf
Interp}^t_n(\Diag_n(\mymod_{m,n}(A),i!))),$$ where ${\sf Interp}_n$
is interpolation at $0,\dots,n-1$. Using algorithms for transpose
interpolation~\cite{KaLa88,BoLeSc03}, this operation can be done in
time $O(\M(n)\log(n))$.  \foorp


\subsection{Associativity Rules}\label{ssec:ar}

For each basic power series operation in Table~\ref{tab:leftcomp}, we
now express $\Eval_{m,n}(A,\o(g))$ in terms of simpler operations; we
call these descriptions \emph{associativity rules}. We write them in a
formal manner: this formalism is the key to automatically design
complex composition algorithms, and makes it straightforward to obtain
\emph{transposed} associativity rules, required in the next section.
Most of these rules are straightforward; care has to be taken
regarding truncation, though.

\smallskip\noindent {\bf Scaling, Shift and Powering.}
\numberwithin{equation}{section}
\begin{gather}
\label{AR:1}\tag{A$_1$}\Eval_{m,n}(A,\M_\lambda(g)) =
\Eval_{m,n}(\Scale_{\lambda,m}(A),g),\\
\label{AR:3}\tag{A$_2$}\Eval_{m,n}(A,\shift_a(g)) = \Eval_{m,n}(\Shift_{a,m}(A),g),\\
\label{AR:4}\tag{A$_3$}
\Eval_{m,n}(A,\power_k(g)) = \Eval_{k(m-1)+1,n}(\Power_{m,k}(A),g).
\end{gather}

%
%

\noindent {\bf Inversion.}  From $A(1/g) =
(\Rev_m(A))(g)/g^{m-1}$ and writing $h = g^{1-m} \bmod x^n$, we get
\begin{equation*}\label{AR:2}\tag{A$_4$}
\Eval_{m,n}(A,\inv(g))=\mul_{n,n}(\Eval_{m,n}(\Rev_m(A),g), h),
\end{equation*}

\smallskip\noindent {\bf Root taking.} For $g$ and $h$ in $\K[[x]]$,
if $g=h^k$, one has $A(h) = A_{0/k}(g) + A_{1/k}(g)h + \cdots
+A_{k-1/k}(g)h^{k-1}$. We deduce the following rule, where the
indices~$m_i$ are defined in Equation~\eqref{eq:ni}.
\begin{equation*}\label{AR:5}\tag{A$_5$}
\hspace{-3.4cm}\begin{array}{l}
h_i = h^i \bmod x^n \text{~for~} 0 \le i < k\\[1mm]
A_0,\dots,A_{k-1} = \Root_{m,k}(A)\notag\\[1mm]
B_i = \Eval_{m_i,n}(A_{i},g) \text{~for~} 0 \le i < k
\end{array}
\end{equation*}
\vskip-4mm
\begin{equation*}
\Eval_{m,n}(A,\root_{k,\alpha,r}(g)) = 
\LinComb_{n}(B_0,\dots,B_{k-1},1,\dots,h_{k-1}).
\end{equation*}

\noindent {\bf Exponential and Logarithm.}
\begin{align}
\label{AR:6}
\Eval_{m,n}(A,\xp(g))& = \Eval_{n}(\Xp_{m,n}(A),g),\tag{A$_6$}\\
\label{AR:7}\tag{A$_7$}
\Eval_{m,n}(A,\lg(g)) &= \Eval_{n}(\Lg_{m,n}(A),g).
\end{align}

%


\subsection{Composition sequences}\label{ssec:comp}
We now describe more complex evaluations schemes, obtained by
composing the former basic ones. 
\begin{Def}Let $\op$ be the set of actions
from Table~\ref{tab:leftcomp}. A sequence $\o=(\o_1,\dots,\o_L)$ with
entries in $\op$ is \emph{defined at a series} $g\in\K[[x]]$ if $g$ is
in $\dom(\o_1)$, and for $i \le L$, $\o_{i-1}(\cdots \o_1(g))$ is in
$\dom(\o_i)$. It is a \emph{composition sequence} if it is defined at
$x$; in this case, $\o$ \emph{computes} the power series
$g_1,\dots,g_L$, with $g_0 = x$ and $g_i = \o_i(g_{i-1})$; it
\emph{outputs} $g_L$.
\end{Def}

\smallskip\noindent{\bf Examples.}  As mentioned in~\cite{Pan98}, the
rational series $g=(ax+b)/(cx+d) \in \K[[x]]$, with $cd \ne 0$,
decomposes as
$$\frac{ax+b}{cx+d} = \frac e{cx+d}+f \text{~with~} e=b-\frac{ad}c\
\text{~and~} f =\frac ac.$$ This shows that $g$ is output by the
composition sequence $(\scale_c, \shift_d, \inv, \scale_e, \shift_f)$.
A more complex example is 
\[g=\frac{2x}{(1+x)^2}=\frac12 \left( 1- \left(1- \frac{2}{1+x} \right)^2 \right),\]
which shows that $g$ is output by the composition
sequence
$$(\shift_1,\inv,\scale_{-2},\shift_{1}, \power_2,\scale_{-1},\shift_1,\scale_{1/2}).$$
Finally, consider 
$g=\log((1+x)/(1-x))$. Using 
$$g= \log \left ( 1 + \left ( -2 -\frac{2}{x-1} \right ) \right ),$$
we get the composition sequence
$(\shift_{-1},\inv,\scale_{-2},\shift_{-2}, \lg).$

\smallskip\noindent{\bf Computing the associated power series.} Our
main algorithm requires truncations of the series $g_1,\dots,g_L$
associated to a composition sequence. The next lemma discusses the
cost of their computation. In all complexity estimates, \emph{the
composition sequence $\o$ is fixed}; hence, our estimates hide a
dependency in $\o$ in their constant factors.

\begin{Lemma}\label{Lemma:0}
  If $\o=(\o_1,\dots,\o_L)$ is a composition sequence that computes
  power series $g_1,\dots,g_L$, one can compute all $g_i \bmod x^n$ in
  time $O(\M(n))$.
\end{Lemma}
\myproof All operators in $\op$ preserve the precision, except for
root-taking, since the operator $\root_{k,\alpha,r}$ loses $r(k-1)$
terms of precision. For $i \le L$, define $\varepsilon_i = r(k-1)$ if
$\o_i$ has the form $\root_{k,\alpha,r}$, $\varepsilon_i = 0$
otherwise, and define $n_L=n$ and inductively $n_{i-1} = n_i +
\varepsilon_i$. Starting the computations with $g_0=x$, we iteratively
compute $g_{i} \bmod x^{n_{i}}$ from $g_{i-1} \bmod x^{n_{i-1}}$.

Inspecting the list of possible cases, one sees that computing $g_i$
always takes time $O(\M(n_{i-1}))$. For powering, this estimate is
valid because we disregard the dependency in $\o$: otherwise, terms of
the form $\log(k)$ would appear. For the same reason, $O(\M(n_{i-1}))$
is in $O(\M(n))$, as is the total cost, obtained by summing over all
$i$. \foorp

\smallskip\noindent{\bf Composition using composition sequences.}  We
now study the cost of computing the map $\Eval_{n}(.,g)$, assuming that
$g \in \K[[x]]$ is output by a composition sequence $\o$. The cost
depends on the operations in $\o$. To keep simple expressions, we
distinguish two cases: if $\o$ contains no operation
$\xp$ or $\lg$, we let $\T_\o(n)=\M(n)$; otherwise, $\T_\o(n)=\M(n)\log(n)$.

\begin{theorem}[Composition]\label{Prop:0}
 Let $\o=(\o_1,\dots,\o_L)$ be a composition sequence that outputs a
 series $g \in \K[[x]]$. Given $\o$, one can compute the map
 $\Eval_{n}(.,g)$ in time $O(\T_\o(n))$.
\end{theorem}
\myproof We follow the algorithm of Figure~\ref{Fig:1}. The main
function first computes the sequence $\g=g_1,\dots,g_L$ modulo $x^n$,
using a subroutine ${\sf ComputeG}(\o, n)$ that follows
Lemma~\ref{Lemma:0}. The cost $O(\M(n))$ of this operation is in
$O(\T_\o(n))$. Then, we call the auxiliary $\Eval{\sf \_aux}$
function.

\begin{figure}[t]
\begin{center}
\fbox{
\begin{minipage}{5 cm}
\begin{tabbing}
\= \quad\quad\quad\quad\quad\quad\= \quad \= \quad \kill
$\underline{\Eval{\sf \_aux}(A, m, n, \ell, \o, \g)}$\\[1mm]
\> {\sf if} $\ell=0$ {\sf return} $A \bmod x^n$\\
\> $\ell'=\ell-1$\\
\> {\sf switch}($\o_\ell$)\\
\> {\sf case}$(\scale_\lambda)$:\> $B = \Scale_{\lambda,m}(A)$\\
\> \> {\sf return} $\Eval{\sf \_aux}(B, m, n, \ell', \o, \g)$\\
\> {\sf case}$(\shift_a)$:\> $B = \Shift_{a,m}(A)$\\
\> \> {\sf return} $\Eval{\sf \_aux}(B, m, n, \ell', \o, \g)$\\
\> {\sf case}$(\power_k)$:\> $B = \Power_{m,k}(A)$\\
\> \> {\sf return} $\Eval{\sf \_aux}(B, km-k+1, n, \ell', \o, \g)$ \\
\> {\sf case}$(\inv)$:\> $B = \Rev_m(A)$\\
\> \> $C = \Eval{\sf \_aux}(B, m, n, \ell', \o, \g)$\\
\> \> {\sf return} $\mul_{n,n}(C,g_{\ell'}^{1-m} \bmod x^n)$\\
\> {\sf case}$(\root_{k,\alpha,r})$:\> $m_0,\dots,m_{k-1} = {\sf FindDegrees}(m, k)$\\
\> \> $h_0=1$\\
\> \> {\sf for} $i=1,\dots,k-1$ {\sf do}\\
\> \> \> $h_i = h h_{i-1} \bmod x^n$\\
\> \> $A_0,\dots,A_{k-1} = {\sf Split}_{m,k}(A)$\\
\> \> {\sf for} $i=0,\dots,k-1$ {\sf do}\\
\> \> \> $B_i = \Eval{\sf \_aux}(A_i, m_i, n, \ell', \o, \g)$\\
\> \> {\sf return} $\LinComb_{n}(B_0,\dots,B_{k-1},h_0,\dots,h_{k-1})$\\
\> {\sf case}$(\xp)$:\> $B = \Xp_{m,n}(A)$\\
\> \> {\sf return} $\Eval{\sf \_aux}(B, n, n, \ell', \o, \g)$\\
\> {\sf case}$(\lg)$:\> $B = \Lg_{m,n}(A)$\\
\> \> {\sf return} $\Eval{\sf \_aux}(B, n, n, \ell', \o, \g)$
\end{tabbing}
\end{minipage}
}

\fbox{
\begin{minipage}{5 cm}
\begin{tabbing}
\= \quad\quad \= \quad \= \quad \kill
$\underline{\Eval(A, n, \o)}$\\[1mm]
\> $\g = {\sf ComputeG}(\o, n)$ \\
\> {\sf return} \Eval{\sf \_aux}$(A, n, n, L, \o, \g)$
\end{tabbing}
\end{minipage}
}
\end{center}
\caption{Algorithm $\Eval$.}
\label{Fig:1}
\end{figure}

On input $A,m,n,\ell,\o,\g$, this latter function computes
$\Eval_{m,n}(A,g_\ell)$. This is done recursively, applying the
appropriate associativity rule~\eqref{AR:1} to~\eqref{AR:7}; the
pseudo-code uses a C-like \texttt{switch} construct to find the
matching case. Even if the initial polynomial $A$ is in $\K[x]_n$,
this may not be the case for the arguments passed to the next calls;
hence the need for the extra parameter $m$. For root-taking, the
subroutine \textsf{FindDegrees} computes the quantities $m_i$ of
Eq.~\eqref{eq:ni}.

Since we write the complexity as a function of $n$, the cost analysis
is simple: even if several recursive calls are generated ($k$ for
$k$th root-taking), their total number is still $O(1)$. Similarly, the
degree of the argument $A$ passed through the recursive calls may
grow, but only like $O(n)$. 

Two kinds of operations contribute to the cost: precomputations of
$g_{\ell-1}^{1-m} \bmod x^n$ (for $\inv$) or of
$1,g_\ell,\dots,g_\ell^{k-1} \bmod x^n$ for $\root_{k,\alpha,r}$, and
linear operations on $A$: shifting, scaling, multiplication \dots The
former take $O(\M(n))$, since the exponents involved are in
$O(n)$. The latter operations take $O(\M(n))$ if no $\Xp$ or $\Lg$
operation is performed, and $O(\M(n)\log(n))$ otherwise. This
concludes the proof. \foorp


\subsection{Inverse map}  

The map $\Eval_n(.,g)$ is invertible if and only if $g'(0) \ne 0$
(hereafter, $g'$ is the derivative of $g$). We discuss here the
computation of the inverse map.
\begin{theorem}[Inverse]\label{Prop:1}x
  Let $\o=(\o_1,\dots,\o_L)$ be a composition sequence that outputs $g
 \in \K[[x]]$ with $g'(0)\ne 0$. One can compute the map
 $\Eval_{n}^{-1}(.,g)$ in time $O(\T_\o(n))$.
\end{theorem}
\myproof If $h$ is the power series $h=\sum_{i \ge i_0} h_i x^{i},$
with $h_{i_0} \ne 0$, $\val(h)=i_0$ is the \emph{valuation} of $h$,
$\ct(h)=h_{i_0}$ its \emph{leading coefficient} and $\tt(h)=h_{i_0}
x^{i_0}$ its \emph{leading term}.  We also introduce an equivalence
relation on power series: $g\sim h$ if $g(0)=h(0)$
and~$\tt(g-g(0))=\tt(h-h(0))$. The proof of the next lemma is
immediate by case inspection.
\begin{Lemma}For $\o$ in
$\op$, if $h \sim g$ and $g$ is in $\dom(\o)$, then $h$ is in
$\dom(\o)$ and $\o(h) \sim \o(g)$.
\end{Lemma}

\smallskip\noindent{\bf Series tangent to the identity.}  We prove the
proposition in two steps. For series of the form $g = x \bmod x^2$,
it suffices to ``reverse'' step-by-step the computation sequence for
$g$. The following lemma is crucial.
\begin{Lemma}\label{Lemma:val}
  Let $g$ be in $\K[[x]]$, with $g = x \bmod x^2$, and let
  $\o=(\o_1,\dots,\o_L)$ be a sequence defined at $g$. Then $\o$ is a
  composition sequence.
\end{Lemma}
\myproof We have to prove that $\o$ is defined at $x$, \emph{i.e.},
that all of $\o_1(x), \o_2(\o_1(x)),\dots$ are well-defined. This
follows by applying the previous lemma inductively. \foorp

\smallskip\noindent We can now work on the inversion property
proper. Let thus $\o=(\o_1,\dots,\o_L)$ be a computation sequence,
that computes $g_1,\dots,g_L$ and outputs $g=g_L$, with $g=x \bmod
x^2$. We define operations $\tilde \o_1,\dots,\tilde \o_L$ through the following
table (note that we reverse the order of the operations):
$$\begin{array}{ccc}
\text{operation} & \o_i & \tilde \o_{L+1-i} \\
\hline 
\text{Add} & \shift_a & \shift_{-a} \\
\text{Mul} & \scale_{\lambda} & \scale_{1/\lambda} \\
\text{Powering} & \power_k & \root_{k,\ct(g_{i-1}),\val(g_{i-1})} \\
\text{Root} & \root_{k,\alpha,r} & \power_k \\
\text{Inverse} & \inv & \inv \\
\text{Exp.} & \xp & \lg \\
\text{Log.} & \lg & \xp \\
\end{array}$$

\begin{Lemma}
 The sequence $\tilde \o=(\tilde \o_1,\dots,\tilde \o_L)$ is a
 composition sequence and outputs a series $\tilde g$ such that
 $\tilde g(g)=x$.
\end{Lemma}
\myproof One sees by induction that for all $i$, $\tilde
\o_{i-1}(\cdots \tilde \o_1(g))$ is in $\dom(\tilde \o_i)$ and $\tilde
\o_{i}(\cdots \tilde \o_1(g))=g_{L-i}$. This shows that the sequence
$\tilde \o$ is defined at $g$ and that $\tilde \o_L(\cdots \tilde
\o_1(g)) = x$.  From Lemma~\ref{Lemma:val}, we deduce that $\tilde \o$
is defined at $x$.  Letting $\tilde g$ be the output of $\tilde \o$,
the previous equality gives $\tilde g(g)=x$, which concludes the
proof. \foorp

\smallskip\noindent Since $\T_\o=\T_{\tilde \o}$, and in view of
Theorem~\ref{Prop:0}, the next lemma concludes the proof of
Theorem~\ref{Prop:1} in the current case.
\begin{Lemma}\label{lemma:rev}
With $g$ and $\tilde g$ as above, the map $\Eval_{n}(.,\tilde g)$ is
the inverse of $\Eval_{n}(.,g)$.
\end{Lemma}
\myproof Let $F$ be in $\K[x]_n$ and let $G=\Eval_{n}(F,g)$, so that
$F(g) = G + H$, with $\val(H)\ge n$. Evaluating at $\tilde g$, we get
$F=G(\tilde g) + H(\tilde g)=G(\tilde g)\bmod x^n$, since $\val(\tilde
g)=1$.  \foorp

\smallskip\noindent{\bf General case.} Lemma~\ref{lemma:rev} fails
when $\val(g)=0$. We can however reduce the general case to that where
$\val(g)=1$. Let us write $g=g_0 +g_1 x +\cdots$, with $g_1 \ne 0$,
and define $\tilde g = (g-g_0)/g_1$, so that $\tilde g=x\bmod x^2$.
If $\o$ is a composition sequence for $g$, then $\tilde \o=(\o,
\shift_{-g_0}, \scale_{1/g_1})$ is a composition sequence for $\tilde
g$, and we have $\T_{\tilde \o} = \T_\o$. Thus, by the previous point,
we can use this composition sequence to compute the map
$\Eval_{n}^{-1}(.,\tilde g)$ in time $O(\T_\o(n))$.  From the equality
$$\Eval_{n}(A,g) = \Eval_{n}(\Scale_{g_1,n}(\Shift_{g_0,n}(A)),\tilde g),$$
we deduce 
$$\Eval_{n}^{-1}(A,g) = \Shift_{-g_0,n}(\Scale_{1/g_1,n}(\Eval_{n}^{-1}(A,\tilde
g))).$$ 
Since scaling and shifting induce only an extra $O(\M(n))$
arithmetic operations, this finishes the proof of Theorem~\ref{Prop:1}.


\section{Change of Basis}  \label{sec:basis}

This section applies our results on composition to \emph{change of
basis} algorithms, between the monomial basis $(x^i)$ and various
families of polynomials $(P_i)$, with $\deg(P_i)=i$, for which we
reach quasi-linear complexity. As an intermediate step, we present a
bivariate evaluation algorithm.


\subsection{Main Theorem}

Let $\F\in \K[[x,t]]$ be the bivariate power series
$$
\begin{array}{l}
\F=\sum_{i,j \ge 0} F_{i,j} x^i t^j =\sum_{j \ge 0} \xi_j(x) t^j.
\end{array}
$$
Associated with $\F$, we consider the map
$$\begin{array}{ccc}
\hspace{-1mm}\Eval_n(.,\F,t):  (a_0,\dots,a_{n-1}) & \mapsto & \sum_{j < n}\xi_j(x) a_j \bmod x^n.
\end{array}$$ The matrix of this map is $[F_{i,j}]_{i,j < n}$.
The following theorem shows that for a large class of series $\F$, the
operation $\Eval_n(.,\F,t)$ and its inverse can be performed
efficiently. The proof relies on a transposition
argument, given in~\S\ref{ssec:transposition}.
\begin{theorem}[Main theorem]\label{Prop:3}
Let $f,g,h,u,v \in \K[[z]]$ be such that
\begin{itemize}
\item $g$ and $h$ are given by composition sequences $\o_g$ and $\o_h$;
\item $f$, $u$ and $v$ can be computed modulo $z^n$ in time $\T(n)$;
\item $g(0)h(0)=0$ and $g'(0),h'(0),u(0),v(0)$ are non-zero;
\item  all coefficients of $f$
  are non-zero.
\end{itemize}
Then the series
$\F(x,t)=u(x)\, v(t)\, f\big (g(x)h(t) \big )$ is
well-defi\-ned. Besides, one can compute the map $\Eval_n(.,\F,t)$ and
its inverse in time $O(\T(n)+\T_{\o_g}(n)+\T_{\o_h}(n))$.
\end{theorem}
\myproof Write $f=\sum_{k \ge 0} f_k z^k$,
$$g(x)^k=\sum_{i \ge 0}
g_{k,i} x^i \quad\text{and}\quad h(t)^k=\sum_{j \ge 0} h_{k,j} t^j.$$ Since
$g(0)h(0)=0$, we have that either
$h_{k,j}=0$ for $k > j$, or $g_{k,i}=0$ for $k>i$. 
Thus, the coefficient
$F^\star_{i,j}$ of $\F^\star$ is well-defined and
$$F^\star_{i,j}=\sum_{k \le n} f_k g_{k,i}h_{k,j}.$$ These
coefficients are those of a product of three matrices, the middle one
being diagonal; we deduce the factorization
$$\Eval_n(.,\F^\star,t)=\Eval_n(.,g) \circ \Diag_n(.,f_i) \circ
\Eval^t_n(.,h).$$ The assumptions on $f$, $g$ and $h$ further imply
that the map $\Eval_n(.,\F^\star,t)$ is invertible, of inverse
$$\Eval_n^{-1}(.,\F^\star,t)=\Eval^{-t}_n(.,h) \circ
\Diag_n(.,f_i^{-1}) \circ \Eval_n^{-1}(.,g).$$ By
Theorems~\ref{Prop:0} and~\ref{Prop:1}, as well as
Theorem~\ref{Prop:4} stated below, $\Eval_n(.,\F^\star,t)$ and its
inverse can thus be evaluated in time
$O(\T(n)+\T_{\o_g}(n)+\T_{\o_h}(n))$. Now, from the identity $\F=u(x)
v(t) \F^\star$, we deduce that
$$\Eval_n(.,\F,t) = \mul_{n,n}(., u) \circ \Eval_n(.,\F^\star,t) \circ
\mul_{n,n}^t(., v).$$ Our assumptions on $u$ and $v$ make this
map invertible, and
$$\Eval_n^{-1}(.,\F,t) = \mul_{n,n}^t(.,b) \circ
\Eval_n^{-1}(.,\F^\star,t) \circ \mul_{n,n}(., a),$$ 
with $a(x)=1/u \bmod x^n$ and $b(t)=1/v \bmod t^n$. The extra
costs induced by the computation of $u$, $v$, their inverses, and
the truncated products fit in $O(\T(n)+\M(n))$.
\foorp


\subsection{Change of Basis}

To conclude, we consider polynomials $(P_i)_{i \ge 0}$ in $\K[x]$,
with $\deg(P_i)=i$, with generating series defined in terms of series $u,v,f,g,h$ as in
Theorem~\ref{Prop:3} by
$${\bf P} = \sum_{i \ge 0} P_i(x) t^i =u(x)\, v(t)\,
f\big(g(x)h(t)\big).$$ 
\begin{Coro}
   Under the above assumptions, one can \sloppy perform the change of basis
   from $(x^i)_{i \ge 0}$ to $(P_i)_{i \ge 0}$, and conversely, in
   time $O(\T(n)+\T_{\o_g}(n)+\T_{\o_h}(n))$.
\end{Coro}
A surprisingly large amount of classical polynomials fits into this
framework (see next section). An important special case is provided
by Sheffer sequences~\cite[Chap.~2]{Roman05},
whose exponential generating function has the form
$$\sum_{i \geq 0} \frac{P_i(x)}{i!}  t^i = v(t) e^{x h(t)}.$$ Examples
include the actuarial, Laguerre, Meixner and Pois\-son-Charlier
polynomials, and the Bernoulli polynomials of the second kind (see
Tables~\ref{Fig:M1} and \ref{Fig:M2}). In this case, if $h$ is output
by the composition sequence $\o$ and $v(t)$ can be computed modulo
$t^n$ in time $\T(n)$, one can perform the change of basis from
$(x^i)_{i \ge 0}$ to $(P_i)_{i \ge 0}$, and conversely, in time
$O(\T(n)+\T_{\o}(n))$.




\subsection{Transposed evaluation}   \label{ssec:transposition}

The following completes the proof of Theorem~\ref{Prop:3}.
\begin{theorem}[Transposition]\label{Prop:4}
 Let $\o=(\o_1,\dots,\o_L)$ be a composition sequence that outputs $g
 \in \K[[x]]$. Given $\o$, one can compute the map
 $\Eval_{n}^{t}(.,g)$ in time $O(\T_\o(n))$.
\end{theorem}

\begin{figure}[t]
\begin{center}
\fbox{
\begin{minipage}{5 cm}
\begin{tabbing}
\= \quad\quad\quad\quad\quad\quad \= \quad \= \quad \kill
$\underline{\Eval{\sf \_aux}^t(A, m, n, \ell, \o, \g)}$\\[1mm]
\> {\sf if} $\ell=0$ {\sf return} $A \bmod x^m$\\
\> $\ell'=\ell-1$\\
\> {\sf switch}($\o_\ell$)\\
\> {\sf case} $(\scale_\lambda)$:\> $B=\Eval{\sf \_aux}^t(A, m, n, \ell', \o, \g)$\\
\> \> {\sf return} $\Scale_{\lambda,m}(B)$\\
\> {\sf case} $(\shift_a)$:\> $B=\Eval{\sf \_aux}^t(A, m, n, \ell', \o, \g)$\\
\> \> {\sf return} $\Shift^t_{a,m}(B)$\\
\> {\sf case} $(\power_k)$:\> $B=\Eval{\sf \_aux}^t(A, mk-m+1, n, \ell', \o, \g)$ \\
\> \> {\sf return} $\Power^t_{m,k}(B)$\\
\> {\sf case} $(\inv)$:\> $B =\mul^t_{n,n}(A,g_{\ell'}^{1-m} \bmod x^n)$\\
\> \> $C = \Eval{\sf \_aux}^t(B, m, n, \ell', \o, \g)$\\
\> \> {\sf return} $\Rev_{m}(C)$\\
\> {\sf case} $(\root_{k,\alpha,r})$:\> $m_0,\dots,m_{k-1} = {\sf FindDegrees}(m, k)$\\
\> \> $h_0=1$\\
\> \> {\sf for} $i=1,\dots,k-1$ {\sf do}\\
\> \> \> $h_i = h h_{i-1} \bmod x^n$\\
\> \> $A_0,\dots,A_{k-1}=\LinComb^t_{n}(A,h_0,\dots,h_{k-1})$\\
\> \> {\sf for} $i=0,\dots,k-1$ {\sf do}\\
\> \> \> $B_i = \Eval{\sf \_aux}^t(A_i, m_i, n, \ell', \o, \g)$\\
\> \> {\sf return} ${\sf Split}^t_{m,k}(B_0,\dots,B_{k-1})$\\
\> {\sf case} $(\xp)$:\> $B=\Eval{\sf \_aux}^t(A, n, n, \ell', \o, \g)$\\
\> \> {\sf return} $\Xp^t_{m,n}(B)$\\
\> {\sf case} $(\lg)$:\> $B=\Eval{\sf \_aux}^t(A, n, n, \ell', \o, \g)$\\
\> \> {\sf return} $\Lg^t_{m,n}(B)$
\end{tabbing}
\end{minipage}
}


\fbox{
\begin{minipage}{5 cm}
\begin{tabbing}
\= \quad\quad \= \quad \= \quad \kill
$\underline{\Eval{\sf Main}^t(A, n, \o)}$\\[1mm]
\> $\g = {\sf ComputeG}(\o, n)$ \\
\> {\sf return} $\Eval{\sf \_aux}^t(A, n, n, L, \o, \g)$
\end{tabbing}
\end{minipage}
}
\end{center}
\caption{Algorithm $\Eval^t$.}
\label{Fig:2}
\end{figure}

\myproof This result follows directly from the transposition
principle. However, we give an explicit construction of the transposed
map $\Eval^t_{n}(.,g)$ in Figure~\ref{Fig:2}. Non-linear
precomputations are left unchanged. The terminal case $\ell=0$ is
dealt with by noting that the transpose of $\mymod_{m,n}$ is
$\mymod_{n,m}$.  To conclude, it suffices to give transposed
associativity rules for our basic operators. The formal approach we
use to write our algorithms pays off now, as it makes this
transposition process automatic.

Recall that our algorithms deal with polynomials. The dual of
$\K[x]_m$ can be identified with $\K[x]_m$ itself: to a $\K$-linear
form $\ell$ over $\K[x]_m$, one associates $\sum_{i <
m}\ell(x^i)x^i$. Hence, transposed versions of algorithms acting on
polynomials are seen to act on polynomials as well.  Remark also that
diagonal operators are their own transpose.

\smallskip\noindent{\bf Multiplication.} In~\cite{BoLeSc03},
following~\cite{HaQuZi04}, details of the transposed versions of
plain, Karatsuba and FFT multiplications are given, with a cost
matching that of the direct product. Without relying on such
techniques, by writing down the multiplication matrix, one sees that
$\mul^t_{n,m}(.,P)$ is
$$A \in \K[x]_m \mapsto (A \Rev_{d+1}(P) \bmod x^{n+d}){\rm~div~}x^{d}
\in \K[x]_n,$$ if $P$ has degree $d$. Using standard multiplication
algorithms, this formulation leads to slower algorithms than those
of~\cite{BoLeSc03}. However, in our usage cases, $n$, $m$ and $d$ are
of the same order of magnitude, and only a constant factor is lost.

\smallskip\noindent {\bf Scale.} The operator $\Scale_{\lambda,n}$ is
diagonal; through transposition, the associativity rule becomes:
\begin{equation*}\label{AR:1t}\tag{A$_1^t$}\Eval^t_{m,n}(A,\M_\lambda(g)) =
\Scale_{\lambda,m}(\Eval^t_{m,n}(A,g)).\end{equation*}

\smallskip\noindent {\bf Shift.} The transposed map $\Rev_n^t$ of the
reversal operator coincides with $\Rev_n$ itself, since this operator
is symmetric. By transposing the identity for
$\Shift$, we deduce
\[\Shift_{a,n}^t(A)=
\Diag_n(\Rev_n(\mul^t_{n,n}(\Rev_n(\Diag_n(A,1/i!)),P)),i!).\]
This algorithm for the transpose operation, though not described as
such, was already given in~\cite{Gerhard00}. This yields:
\begin{equation*}\label{AR:3t}\tag{A$_2^t$}\Eval_{m,n}^t(A,\shift_a(g)) = \Shift_{a,m}^t(\Eval^t_{m,n}(A,g)).\end{equation*}
                                                      
\smallskip\noindent {\bf Powering.}  The dual map $\Power_{n,k}^t$
maps $A \in \K[x]_{k(n-1)+1}$ to $A_{0/k}\in\K[x]_n$ (with the notation
of \S\ref{ssec:basic}). We deduce:
\begin{equation*}\label{AR:4t}\tag{A$_3^t$}
\Eval_{m,n}^t(A,\power_k(g)) = \Power_{m,k}^t(\Eval^t_{k(m-1)+1,n}(A,g)).
\end{equation*}
 
\smallskip\noindent {\bf Inversion.}  The transposed version of the
rule for $\inv$ is
\begin{equation*}\label{AR:2t}\tag{A$_4^t$}
 \Eval_{m,n}^t(A,\inv(g)) =
\Rev_m(\Eval_{m,n}^t(\mul_{n,n}^t(A,g^{1-m}),g)).
\end{equation*}

\smallskip\noindent {\bf Root taking.}
Considering its matrix, one sees that 
$\Root_{m,k}^t$ maps $(A_0,\dots,A_{k-1}) \in \K[x]_{m_0} \times
\cdots \times \K[x]_{m_{k-1}}$ to $$ A_0(x^k) + A_1(x^k) x + \cdots +
A_{k-1}(x^k)x^{k-1} \in \K[x]_m.$$ Besides, since the map $\LinComb$ is
the direct sum of the maps
$$\mul_{n,n}(.,G_i): \K[x]_{n} \to \K[x]_n,$$
its transpose
$\LinComb^t_{n}(.,G_0,\dots,G_{k-1})$ sends $A \in \K[x]_n$ to 
$$ (\mul^t_{n,n}(A,G_i))_{0\le i \le k-1} 
 \in \K[x]_{n}^k.$$ 
Putting this together gives the transposed associativity rule
\begin{equation*}\label{AR:5t}
\begin{array}{l}
  h_i = h^i \bmod x^n \text{~for~} 0 \le i < k\\[1mm]
  A_0,\dots,A_{k-1} = \LinComb^t_{n}(A,h_0,\dots,h_{k-1})\\[1mm]
  B_i = \Eval^t_{m_i,n}(A_{i},g) \text{~for~} 0 \le i < k\\[1mm] 
 \Eval^t_{m,n}(A,\root_{k,\alpha,r}(g)) = \Root^t_{m,k}(B_0,\dots,B_{k-1})\notag\\[1mm]
\end{array}
\tag{A$_5^t$}
\end{equation*}

\noindent{\bf Exponential and Logarithm.}  From the proof of
Proposition~\ref{Prop:xp}, we deduce the transposed map of $\Xp_{m,n},
\Lg_{m,n}$ and their associativity rules
\begin{gather}
\Xp^t_{m,n}(A) = \mymod_{n,m} (\Shift_{-1,n}^t({\sf MultiEval}(\Diag_n(A,1/i!)))),\notag\\
\label{AR:6t}\Eval^t_{m,n}(A,\xp(g))= \Xp^t_{m,n}(\Eval^t_{n,n}(A,g));\tag{A$_6^t$}\\
\Lg^t_{m,n}(A) = \mymod_{n,m} (\Diag_n({\sf Interp}_n(\Shift_{1,n}^t(A)),i!)),\notag\\
\label{AR:7t}
\Eval^t_{m,n}(A,\lg(g)) = \Lg^t_{m,n}(\Eval^t_{n,n}(A,g)).\tag{A$_7^t$}
\end{gather}


\section{Applications}\label{sec:applications}

Many generating functions of classical families of polynomials fit
into our framework. To obtain conversion algorithms, it is sufficient
to find suitable composition sequences. Table~\ref{Fig:M1} lists
families of polynomials for which conversions can be done in time
$O(\M(n))$ with our method (see e.g.~\cite{Roman05,AnAsRo99} for more
on these classical families). In Table~\ref{Fig:M2}, a similar list is
given, leading to conversions of cost~$O(\M(n)\log n)$; most of these
entries are actually Sheffer sequences. Many other families can be
obtained as special cases (e.g., Gegenbauer, Legendre, Chebyshev,
Mittag-Leffler, etc). 

The entry marked by $(\star)$ is from~\cite{Gerhard00}; the entries
marked by $(\star\star)$ are orthogonal polynomials, for which one
conversion (from the orthogonal to the monomial basis) is already
mentioned with the same complexity in~\cite{DrHeRo97,PoStTa98}.

In all cases, the pre-multiplier $u(x)v(t)$ depends on $t$ only and
can be computed at precision $n$ in time $O(\M(n))$; all our functions
$f$ can be expanded at precision $n$ in time $O(n)$. Regarding the
functions $g(x)$ and $h(t)$, most entries are easy to check; the only
explanations needed concern some series $h(t)$. Rational functions are
covered by the first example of \S\ref{ssec:comp}; the second example
of \S\ref{ssec:comp} deals with Jacobi polynomials and Spread
polynomials; the last example of \S\ref{ssec:comp} shows how to handle
functions with logarithms.  For Fibonacci polynomials, the function
$h(t)=t/(1-t^2)$ satisfies  $$(2h)^2 = \Big
(\frac{1+t^2}{1-t^2}\Big )^2-1.$$ From this, we deduce the sequence
for $h$:
$$(\power_2, \scale_{-1}, \shift_1, \inv, \scale_2, \shift_{-1},
 \power_2, \shift_{-1}, \root_{2,2,1}, \scale_{1/2}).$$ For
 Mott polynomials the series $h(t)=(1-\sqrt{1-t^2})/t$ can be
 rewritten as
$$h=\sqrt{\frac{2}{1+\sqrt{1-t^2}}-1}.$$ This yields the composition
sequence
$$(\power_2, \scale_{-1}, \shift_1, \root_{2,1,0}, \shift_1, \inv, \scale_2, \shift_{-1}, \root_{2,1,0}).$$

\begin{table*}[!!!t]
$$
\begin{array}{l|c|c|c|c|c}
\text{polynomial} &\text{generating series} & u(x)v(t) & f(z) & g(x) & h(t) \\
\hline
\text{Laguerre}\ L_n^\alpha & {\sum_{n \ge 0} L_n^{\alpha}(x) t^n} &
(1-t)^{-1-\alpha} & \exp(z) & -x & t(1-t)^{-1}\\

\text{Hermite}\ H_n & \sum_{n \ge 0} \frac{1}{n!} H_n(x) t^n &
{\exp(-t^2)}& \exp(z) & 2x & t\\

\text{Jacobi}\ P_n^{(\alpha,\beta)}&
\sum_{n \ge 0}\frac{(\alpha+\beta+1)_n }{(\beta+1)_n} P_n^{(\alpha,\beta)}(x) t^n&
{{(1+t)^{-\alpha-\beta-1}}}&
{{{}_2F_1}(\frac{\alpha+\beta+1}2,\frac{\alpha+\beta+2}2;\beta+1;z)}
& 1+x &{2t{(1+t)^{-2}}}\\

\text{Fibonacci}\ F_n & \sum_{n \ge 0} F_n(x) t^n &
{{(1-t^2)^{-1}}}&{{(1 - z)^{-1}}} & x &
{{t}{(1-t^2)^{-1}}}\\

\text{Euler}\  E_n^\alpha & \sum_{n \ge 0} \frac{1}{n!} E_n^\alpha(x)t^n &
{2^\alpha ({e^t+1})^{-\alpha}}&
\exp(z) & x & t \\

\text{Bernoulli}\  B_n^\alpha &  \sum_{n \ge 0} \frac{1}{n!}B_n^\alpha(x)t^n &
t^\alpha(e^t-1)^{-\alpha}&
\exp(z) & x & t \\

\text{Mott}\ M_n & \sum_{n \ge 0} \frac{1}{n!} M_n(x)t^n &
1 & \exp(z) & -x &  {({1-\sqrt{1-t^2}})/t}  \\

\text{Spread}\ S_n & \sum_{n \ge 0} S_n(x) t^n &
{({1+t})({1-t})^{-1}}&{{z}({1+4z})^{-1}}
&  x &{t{(1-t)^{-2}}}\\

\text{Bessel}\ p_n & \sum_{n \ge 0} \frac{1}{n!} p_n(x)t^n &
1 &\exp(z) & x & 1-\sqrt{1-2t}
\end{array}
$$
\vskip-3ex
\caption{Polynomials with conversion in $O(\M(n))$}\label{Fig:M1}

\vskip-1ex
$$\begin{array}{l|c|c|c|c|c}
\text{polynomial} &\text{generating series} & u(x)v(t) & f(z) & g(x) & h(t) \\
\hline
\text{Falling factorial}\ (x)_n \hfill (\star) & \sum_{n \ge 0} \frac{1}{n!} (x)_n t^n
 & 1 & \exp(z) & x & \log(1+t) \\

\text{Bell}\ \phi_n & \sum_{n \ge 0} \frac{1}{n!} \phi_n(x) t^n
 & 1 & \exp(z) & x & \exp(t)-1 \\

\text{Bernoulli, 2nd kind}\ b_n & \sum_{n \ge 0} \frac{1}{n!} b_n(x)t^n &{{t}/{\log(1+t)}} &
\exp(z) & x & \log(1+t)\\

\text{Poisson-Charlier}\ c_n(x;a) &  \sum_{n \ge 0} \frac{1}{n!} c_n(x;a)t^n & \exp(-t) & \exp(z) & x & \log(1+t/a)\\

\text{Actuarial}\ a_n^{(\beta)} &  \sum_{n \ge 0} \frac{1}{n!} a_n^{(\beta)}(x) t^n & \exp(\beta t) &
\exp(z) & -x & \exp(t)-1\\

\text{Narumi}\ N^{(a)}_n & \sum_{n \ge 0} \frac{1}{n!} N^{(a)}_n(x) t^n & {t}^a{\log(1+t)}^{-a}
& \exp(z) & x & \log(1+t)\\

\text{Peters}\ P^{(\lambda,\mu)}_n & \sum_{n \ge 0} \frac{1}{n!} P^{(\lambda,\mu)}_n(x)t^n &  {(1+(1+t)^\lambda)^{-\mu}}& \exp(z) & x& \log(1+t)\\

\text{Meixner-Pollaczek}\  P_n^{(\lambda)}(x;\phi){~~} (\star\star)&  \sum_{n \ge 0} P_n^{(\lambda)}(x;\phi) t^n &
(1+t^2-2 t \cos \phi)^{-\lambda}
& \exp(z) & ix & \log ( \frac{1-t e^{i \phi}}{1-t e^{-i \phi}})\\

\text{Meixner}\ m_n(x;\beta,c) \hfill (\star\star)&  \sum_{n \ge 0} \frac{(\beta)_n}{n!} m_n(x;\beta,c) t^n
& {(1-t)^{-\beta}} & \exp(z) & x & \log ( \frac{1-t/c}{1-t})\\

\text{Krawtchouk}\ K_n(x;p,N) \hfill (\star\star)&  \sum_{n \ge 0} \binom{N}{n} K_n(x;p,N) t^n
& (1+t)^N & \exp(z) & x & \log ( \frac{p-(1-p)t}{p(1+t)})\\
\end{array}
$$
\vskip-3ex\caption{Polynomials with conversion in $O(\M(n)\log(n))$}\label{Fig:M2}
\end{table*}



\section{Experiments}

We implemented the algorithms for change of basis using
NTL~\cite{Shoup95}; the experiments are done for coefficients defined
modulo a 40 bit prime, using the \verb+ZZ_p+ NTL class (our algorithms
still work for degrees small with respect to the characteristic). All
timings reported here are obtained on a Pentium M, 1.73 Ghz, with 1 GB
memory.

Our implementation follows directly the presentation of the former
sections. We use the transposed multiplication implementation
of~\cite{BoLeSc03}. The Newton iteration for inverse is built-in in
NTL; we use the standard Newton iteration for square
root~\cite{Brent75}. Exponentials are computed using the algorithm
of~\cite{HaZi04}. Powers are computed through exponential and
logarithm~\cite{Brent75}, except when the arguments are binomials,
when faster formulas for binomial series are used. For evaluation and
interpolation at $0,\dots,n-1$, and their transposes, we use the
implementation of~\cite{BoLeSc03}.

\begin{figure}[!!!!!!!!!h]
\centerline{\includegraphics[scale=0.5]{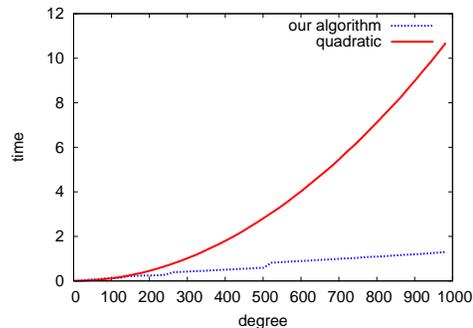}}
\vspace{-3ex}
\caption{{\small Jacobi polynomials.}}
\label{Fig:jacobi}
\end{figure}

\begin{figure}[!!!!!!!!!h]
\centerline{\includegraphics[scale=0.5]{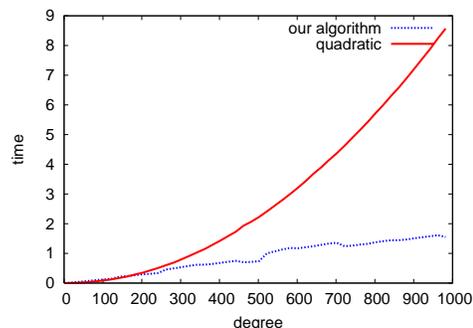}}
\vspace{-3ex}
\caption{{\small Mittag-Leffler polynomials.}}
\label{Fig:mittag}
\end{figure}

We use the Jacobi and Mittag-Leffler orthogonal polynomials (a special
case of Meixner polynomials, with $\beta=0$ and $c=-1$), with the
composition sequences of \S\ref{ssec:comp}. Our algorithm has cost
$O(\M(n))$ for the former and $O(\M(n)\log(n))$ for the latter.  We
compare this to the naive approach of quadratic cost in
Figure~\ref{Fig:jacobi} and~\ref{Fig:mittag}, respectively. Timings
are given for the conversion from orthogonal to monomial bases; those
for the inverse conversion are similar.
\vspace{-5ex}

Our algorithm performs better than the quadratic one. The crossover
points lie between 100 and 200; this large value is due to the
constant hidden in our big-Oh estimates: in both cases, there is a
contribution of about $20\M(n)$, plus an additional $\M(n)\log(n)$ for
Mittag-Leffler.



\section{Discussion}

This article provides a flexible framework for generating new families
of conversion algorithms: it suffices to add new composition operators
to Table~\ref{tab:leftcomp} and provide the corresponding
associativity rules.  Still, several questions need further
investigation. Several of the composition sequences we use are
non-trivial: this raises in particular the questions of characterizing
what functions can be computed by a composition sequence, and of
determining such sequences algorithmically.  Besides, the costs of our
algorithms are measured only in terms of arithmetic operations; the
questions of numerical stability (for floating-point computations) or
of coefficient size (when working over $\Q$) require further work.

\smallskip\noindent{\bf Acknowledgments.} We thank ANR Gecko, the
joint Inria-Microsoft Research Lab and NSERC for financial support.

\scriptsize

\end{document}